\documentstyle[prl,aps,twocolumn,epsfig]{revtex} 
\begin{document} 
\preprint{} 
\draft 

\title{ Measurement-induced Squeezing of  
              a Bose-Einstein Condensate        } 
\author{ D. A. R. Dalvit${}^{1}$, 
         J. Dziarmaga$^{1,2}$, 
         and R. Onofrio${}^{1,3}$} 
\address{ 
${}^1$ Los Alamos National Laboratory, Los Alamos, 
       New Mexico 87545 \\ 
${}^2$ Intytut Fizyki Uniwersytetu Jagiello\'nskiego, 
       30-059 Krak\'ow, Poland \\ 
${}^3$ Dipartimento di Fisica "G. Galilei", 
       Universit\`a di Padova, Padova 35131, Italy \\
       and 
       INFM, Sezione di Roma "La Sapienza", Roma 00185, Italy} 
\maketitle 

\begin{abstract} 

We discuss the dynamics of a Bose-Einstein condensate
during its nondestructive imaging. A generalized Lindblad
superoperator in the condensate master equation is used to
include the effect of the measurement. A continuous
imaging with a sufficiently high laser intensity
progressively drives the quantum state of the condensate into
number squeezed states.
Observable consequences of such a measurement-induced
squeezing are discussed.

\end{abstract} 


  Since its birth, quantum mechanics has led to an interpretational
debate on the role played by the
measurement process in its structure and its relationship
to classical mechanics developed for macroscopic systems
\cite{Wheeler}. This debate has been
enriched by the realization of new experimental techniques 
spanning from quantum jumps in single ion traps to
macroscopic entangled states in various quantum systems. 
Recently, the production of atomic Bose-Einstein condensates 
of dilute atomic gases has also paved the way to the study 
of dynamical phenomena of macroscopic quantum systems with the precision
characteristic of atomic physics \cite{Varenna}. 

Two optical techniques, absorption and dispersive imagings, 
have been used to monitor the dynamics of a Bose-Einstein condensate 
\cite{Varenna}. 
In the former the condensate interacts with a light beam resonant 
(or close to resonance) with an atomic transition.  The output beam is
attenuated proportionally to the column density of the
condensate - the condensate density integrated along the
line of sight of the imaging beam. The absorption of
photons heats the condensate then strongly perturbing it,
and in general a new replica of the condensate has to be
produced to further study its dynamics. This measurement
is, in the language of quantum measurement theory, of type-II
since it destroys the state of the observed system and
forbids the study of the dynamics of a single quantum
system \cite{Pauli}. Repeated measurements on a
Bose-Einstein condensate or, at the limit, its continuous
monitoring are instead possible using its dispersive features, 
for instance through
phase-contrast \cite{Andrews} or interference
\cite{Kadlecek} imaging techniques. Off-resonance light is
scattered by the condensate which induces phase-shifts
thereby converted into light intensity modulations by
homodyne or heterodyne detection.  The off-resonant nature
of the atom-photon interaction allows for a very low
absorption rate and therefore low heating of the
condensate.  Thus, multiple shots of the same condensate
can be taken - a type-I measurement - allowing to study
with high accuracy several phenomena, like its formation in
non-adiabatic conditions \cite{Miesner}, short and long
wavelength collective excitations \cite{Andrews1}, vortices
and superfluid dynamics \cite{Matthews}. The effect of the
measurement process is typically neglected in these
analysis. A first attempt to include the measurement
process in a two-mode configuration for the condensate has
been discussed in \cite{Corney}. 
Our main goal is to
include the atom-photon interaction process present in
dispersive imaging into the intrinsic dynamics of the
condensate. We show that the measurement process induces number 
squeezing of the 
quantum state of the condensate, and we provide realistic estimates that 
suggest that the phenomenon could be observed with current experimental 
techniques.

Let us start the analysis with the effective interaction
Hamiltonian between the off-resonant photons and the atoms,
written as:

\begin{equation} 
\hat{H}_{\rm int}= {{\epsilon_0 \chi_0} \over 2} 
\int d^3x~\hat{n}({\bf x}) : {\bf E}^2 : ,  
\label{Intham} \end{equation} 
where $\hat{n}$ is the condensate density operator, and 
${\bf E}$ the electric field due to the intensity $I$ of the
incoming light.  The coefficient $\chi_0$ represents the
effective electric susceptibility of the atoms defined as
$\chi_0=\lambda^3 \delta/2 \pi^2 (1+\delta^2)$, where we
have introduced the light wavelength $\lambda$ and the
light detuning measured in half-linewidths $\Gamma/2$ of
the atomic transition, $\delta=(\omega-\omega_{\rm
at})/(\Gamma/2)$. Equation (\ref{Intham}) allows us to
write the reduced master equation for the atomic degrees of
freedom by a standard technique, {\it i.e.} by tracing out
the photon degrees of freedom \cite{Carmichael}. The
photons are assumed to be in a plane-wave state with
momentum along the impinging direction, corresponding to a
wavevector ${\bf k}=k_0 \hat{z}$ orthogonal to the imaging
plane $x-y$.  Unless tomographic techniques are used, the
image results from a projection of the condensate onto the
$x-y$ plane, by integrating along the $z$ direction. This
demands to project the dynamics of the condensate into the
imaging plane. In order to write a closed 2-D master
equation to describe the $x-y$ dynamics we assume the
condensate wavefunction to be factorizable as
$\psi(x,y,z)=\phi(x,y) \Lambda(z)$. Such factorization
holds if the confinement in the $z$-direction is strong
enough to make the corresponding mean-field energy
negligible with respect to the energy quanta of the
confinement, {\it i.e.} $\hbar \omega_z >> N g$, 
as recently demonstrated in \cite{lowerdim},
where $g=4 \pi \hbar^2 a/m$, with $a$ the s-wave scattering length,
and $\omega_z$ the angular frequency of the confinement
harmonic potential along the $z$ direction. 
The resulting reduced master equation in the imaging plane is written as
\cite{long}:

\begin{eqnarray} 
{{d \hat\rho} \over dt}&=&
-{i \over \hbar} [\hat{H}_{\rm{2D}}, \hat{\rho}] \nonumber \\
&&
-\int d^2r_1 \int d^2r_2 K({\bf r}_1-{\bf r}_2) 
[\hat{n}({\bf r}_1),[\hat{n}({\bf r}_2),\hat{\rho}]] , 
\label{LINDBLAD} 
\end{eqnarray} 
where $\hat{n}({\bf r})=
\hat{\Psi}^{\dagger}\hat{\Psi}({\bf r})$ is the 2-D density
operator. The effect of the measurement is taken into
account through the second term in the right-hand side of
Eq.(\ref{LINDBLAD}). This equation preserves the total
number of atoms, and corresponds to a quantum
nondemolition \cite{nondemo} coupling between the atom and the optical
fields \cite{Corney,Onofrio1,Li,Leonhardt}. The measurement
kernel $K$ has the expression:

\begin{equation}
K({\bf r})= {{\pi \chi_0^2 k_0 I} \over 2 \hbar c} 
\int d^2k \exp(-\xi^2 k^4/4 k_0^2) \exp(i {\bf k} {\bf r}) ,  
\label{KERNEL}
\end{equation} 
where $\xi$ is the lengthscale of the condensate in the $z$
direction, the width of the Gaussian state $\Lambda(z)$
under the abovementioned approximation. Equation
(\ref{KERNEL}) holds for a condensate having thickness in the $z$ 
direction $\xi \gg \lambda$. If the
measurement kernel were a local one, $K({\bf r}_1-{\bf
r}_2) \simeq \delta({\bf r}_1 -{\bf r}_2)$,
Eq.(\ref{LINDBLAD}) would reduce to a Lindblad equation for
the measurement of an infinite number of densities
$\hat{n}({\bf r})$. This assumes that no spatial
correlation is estabilished by the photon detection.
However, the ultimate resolution limit in the imaging
system depends on the photon wavelength, regardless of the
pixel density of the detecting camera. The resolution
lengthscale follows from Eq.(\ref{KERNEL}) as a width of
the kernel $\Delta r=(2\pi^2\xi/k_0)^{1/2}=(\pi \xi
\lambda)^{1/2}$, the geometrical average of the light
wavelength and the condensate thickness $\xi$.

When the measurement outweighs the self dynamics, the wavefunction
of the condensate is driven towards an eigenstate of the measurement apparatus.
In that case the master equation (\ref{LINDBLAD})
is solved by the
eigenstates of the density operator $\hat{n}({\bf r})$, 
which are then pointer states \cite{ZUREK},
$ \hat{\Psi}^{\dagger}({\bf r}_1)\dots
\hat{\Psi}^{\dagger}({\bf r}_N)|0\rangle$.
This state describes a ``gas'' of $N$ atoms localized at
the points ${\bf r}_1\dots {\bf r}_N$. These states are
manifestly not mean-field states so they cannot be
described by any generalization of the Gross-Pitaevskii equation. A
measurement strong in comparison to the Hamiltonian
$\hat{H}_{\rm{2D}}$ will project an initial mean field
condensate state onto one of the pointer states with a
probability distribution for different ${\bf r}$'s given by
the initial mean field wave function $p({\bf
r})=\phi^{\star}\phi({\bf r})$. However, this dramatic
change of the condensate state will not directly affect the
condensate image since the homodyne current is given by
the convolution $I({\bf r})\sim\int d^2r'K({\bf r}-{\bf
r}') \langle\hat{n}({\bf r}')\rangle$, which coarse-grains
the density distribution over the resolution of the kernel
$\Delta r^2$.  The only effect on the image will be small
statistical fluctuations:  if the average number of atoms
within a given resolution area of $\Delta r^2$ is $n$, then
after the localization the number will be in the range $n\pm\sqrt{n}$.

When the system Hamiltonian $H_{\rm{2D}}$ is strong enough, it
can change pointer states \cite{SH}. 
To quantify this phenomenon and the competition between localization 
induced by the measurement and the delocalization due to 
$\hat{H}_{\rm{2D}}$ we discretize Eq.(\ref{LINDBLAD}) introducing 
a 2-D lattice with the natural choice for the lattice
constant set by the resolution lengthscale $\Delta r$,

\begin{equation} 
{{d \hat\rho} \over dt}=
-\frac{i}{\hbar}
[ -\hbar\omega\sum_{\langle k,l \rangle} 
   \hat{\Psi}^{\dagger}_k \hat{\Psi}_l+\hat{V}, \hat{\rho}\;] 
- S \sum_l [\hat{n}_l,[\hat{n}_l,\hat{\rho}]]. 
\label{LATTICE} 
\end{equation}
Here $\hat{\Psi}_l$ is an annihilation operator and
$\hat{n}_l = \hat{\Psi}_l^{\dagger} \hat{\Psi}_l$ is the
number operator at a lattice site $l$.  The frequency of
hopping between any nearest neighbor sites $\langle k,l
\rangle$ is $\omega\approx \hbar/ 2m \Delta r^2$, which is
$\hbar^{-1}$ times the characteristic kinetic energy.  The
potential energy operator is $\hat{V}=\sum_l
U_l\hat{n}_l+G\hat{n}_l^2$, where $U_l$ is the trapping
potential and $G=g / \xi\Delta r^2$. The effective
measurement strength is 
$S\approx \int d^2r K({\bf r}) / \Delta r^2  = (2\pi /\Delta r)^2 
(\pi \chi_0^2 k_0 I /2\hbar c)$. 

  The Lindblad term in Eq. (\ref{LATTICE}) is a sum
of Lindblad terms for different lattice sites. Its pointer
states are the Fock states $|\{N_l\}\rangle = |N_1, N_2,
\ldots \rangle$ with definite occupation numbers $N_l$ at
every site $l$. An initial pure mean-field state can be
expanded in the Fock basis, $|\psi(0)\rangle=
\sum_{\{N_l\}} \psi_{\{N_l\}}(0) |\{N_l\}\rangle$. If the
initial mean-field wavefunction of the condensate
is $\phi_l$, then the initial expectation value of the
number of atoms at a site $l$ is
$n_l(0)=N\phi^{\star}_l\phi_l$. Each site has a finite
dispersion in the number of atoms,
$\sigma_l(0)\simeq[n_l(0)]^{1/2}$.  For  large $n_l(0)$
the distribution of $N_l$ can be approximated by a
Gaussian. If there were no self-Hamiltonian the Lindblad
term would drive the state of the system into a mixed state
of different Fock states $|\{N_l\}\rangle$. The mixed state
describes an ensemble of outcomes of different realizations
of the experiment. In a given realisation the outcome would
be a definite Fock state $|\{N_l\}\rangle$ with random
$N_l$ in the range $n_l(0)\pm\sigma_l(0)$. To describe a single
realisation of the experiment we use a (Stratonovich)
unraveling of the master equation by a stochastic
Schr\"odinger equation (SSE) \cite{Gatarek}

\begin{eqnarray}
&& \frac{d}{dt} \psi_{\{N_l\}} =
-\frac{i}{\hbar}
\sum_{\{N'_l\}}
h_{\{N_l,N'_l\}}
\psi_{\{N'_l\}}
-\frac{i}{\hbar}
V_{\{N_l\}}
\psi_{\{N_l\}} 
\nonumber \\
&& + \psi_{\{N_l\}} \sum_l
\left[
-S(N_l-n_l)^2+
S\sigma_l^2+
(N_l-n_l)\theta_l
\right]  \;\;,
\label{SSE}
\end{eqnarray} 
where the homodyne noises have averages
$\overline{\theta_{l}(t)}=0$ and
$\overline{\theta_{l_1}(t_1)\theta_{l_2}(t_2)}=
2S\delta_{l_1,l_2}\delta(t_1-t_2)$. Here
$n_l=\sum_{\{N_l\}}N_l|\psi_{\{N_l\}}|^2$, 
$\sigma_l=\sum_{\{N_l\}}(N_l-n_l)^2|\psi_{\{N_l\}}|^2$, 
$h$ is the matrix element of the hopping Hamiltonian and
$V_{\{N_l\}}=\sum_l(U_lN_l+GN_l^2)$ is the potential
energy.  The unravelling does not change the pointer states
\cite{DDZ}.

\begin{figure}[h] 
\centering 
\epsfig{file=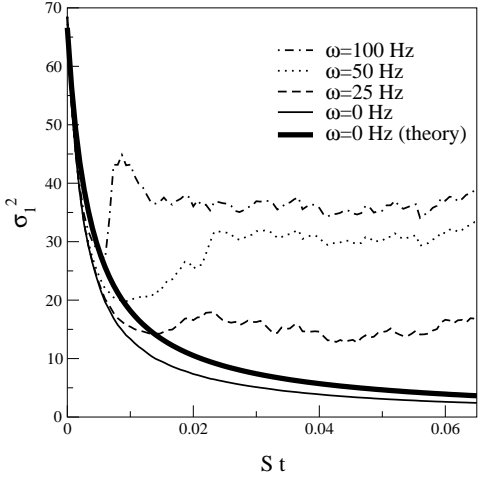, width=7cm, clip=}
\caption{ $\sigma_1^2(t)$ for a 1-D 3-mode configuration  
with $S=2 \rm{s}^{-1}$, $n_l(0)=100$ and $\sigma_l^2(0)=66.6$. We show 
four single realizations of the experiment for different hopping
frequencies. The thick solid line is the analytic solution 
$\sigma_l^2(t)= \sigma_l^2(0)/[1 + 4St\sigma_1^2(0)]$ 
for $\omega=0 \; \rm{s}^{-1}$. }
\label{figure1} 
\end{figure} 
Equation (\ref{SSE}) can be solved provided that we neglect the
hopping term, $h=0$.  In this case Fock states are fixed
points because they are characterized by $N_l=n_l$ and $\sigma_l=0$.
SSE is satisfied {\it exactly} by the ansatz

\begin{equation}
\psi_{\{N_l\}}(t)=
e^{i\varphi_{\{N_l\}}}
e^{ -\frac{i}{\hbar}V_{\{N_l\}} t}
\prod_l \frac{ e^{-\frac{[N_l-n_l(t)]^2}{4\sigma_l^2(t)}} }
             {         [2\pi\sigma_l^2(t)]^{1/4}          } ,
\label{ANSATZ}
\end{equation}
when the parameters satisfy the equations of motion

\begin{equation}
\frac{d}{dt} \sigma_l^2 = -4 S \sigma_l^4,\;\;\;
\frac{d}{dt} n_l = 2 \sigma_l^2 \theta_l ,
\nonumber
\end{equation}
and $\varphi_{\{N_l\}}=\sum_l N_l {\rm Arg}(\phi_l)$ are constant
phases. The ansatz remains normalized as long as
$\sigma_l^2\gg 1$ when summations over $N_l$ can be
replaced by integrals. For an initial $\sigma_l^2(0)\simeq
n_l(0)$ the dispersions shrink to zero (a Fock state) as
$\sigma_l^2(t)=\sigma_l^2(0)/[1+4\sigma_l^2(0)St]$. 
Intuitively, the continuous reduction of number fluctuations is due to
the weak nature of the measurement, and takes place on a time scale
$\simeq 1/ S \sigma^2_l(0)$.
While $\sigma_l^2$'s shrink down to $\sigma_l^2\approx 1$ the
means $n_l$ make random walks in the range
$n_l(0)\pm\sigma_l(0)$. When $\sigma_l^2\ll 1$ our ansatz
breaks down but the numerical simulations for 2- and
3-site models show that $n_l$ becomes frozen.

For non zero $h$, localization in a Fock state may be
inhibited due to the hopping between nearest neighbor
sites. Here a scaling argument is used to estimate the
dependence on the physical parameters of 
$\sigma_l^2(\infty)$, the dispersion at which
the measurement term and the hopping term in the SSE
balance each other.  We assume that the density
distribution is smooth, i.e. any two nearest neighbor sites
$k,l$ have close occupation numbers, $n_k\approx n_l$.  
For large occupation numbers, the element of the hopping
Hamiltonian between $k$ and $l$ can be approximated by
$\omega \sqrt{N_k N_l} \approx \omega\sqrt{n_k n_l} \approx
\omega n_l$.  The hopping mixes the amplitudes
$\psi_{\{N_l\}}$ and $\psi_{\{N'_l\}}$ which have
negligibly close potential energies, $V_{\{N_l\}}\approx
V_{\{N'_l\}}$.  We again assume that $\sigma_l^2\gg 1$ so
that we can treat the occupation numbers $N_l$ as
continuous variables.  After
neglecting the noise term, the SSE can be written as an
eigenvalue ($\Omega$) problem

\begin{eqnarray}
&& 
i\Omega \psi_{ \{N_l\}} =
\psi_{\{N_l\}}
\sum_l \left[ -S(N_l-n_l)^2 + S \sigma_l^2 \right] 
\nonumber\\
&&
-i \omega \sum_{\langle k,l \rangle} 
\sqrt{n_k n_l}
\left[
\frac{ \partial^2 \psi_{\{N_l\}} } { \partial N_k^2 } +
\frac{ \partial^2 \psi_{\{N_l\}} } { \partial N_l^2 } -
2 \frac{ \partial^2 \psi_{\{N_l\}} } { \partial N_k \partial N_l } 
\right]  \;\; .
\label{SCALE}
\end{eqnarray}  
This equation is made parameter independent by rescaling
$N_l\equiv \tilde{N}_l (\omega n_l/S)^{1/4}$ and
$\Omega=\tilde{\Omega}\sum_l (S\omega n_l)^{1/2}$, so the
stationary solution of the rescaled equation must scale as 
$\sigma_l(\infty)\simeq(\omega n_l/S)^{1/4}$. In
Figures 1 and 2 we show results of exact numerical
simulations on a 1-D 3-site periodic lattice with 300 atoms
which confirm the validity of these predictions. The solution
without the hopping term,
$\sigma^2_l(t)=\sigma_l^2(0)/[1+4St\sigma_l^2(0)]$, reaches
its asymptotic value $\sigma_l(\infty)$ after a time 
$\tau_l\simeq[S\omega n_l(0)]^{-1/2}$.

We estimate the effect assuming the following
parameters for the condensate and its imaging, relevant for
the case of $^{87}{\rm Rb}$:  $m=1.4\;10^{-25}{\rm kg}$,
$a=5.8\rm{nm}$, $\lambda=780{\rm nm}$, $\chi_0= 10^{-23}{\rm
m}^3$, laser intensity $I=100 {\rm mW}/{\rm cm}^2$, and a
total number of atoms $N=10^7$. The width of the Gaussian
$\Lambda(z)$ is assumed to be $\xi=10\mu{\rm m}$.  The
resolution of the kernel (\ref{KERNEL}) is $\Delta
r=(\pi\lambda\xi)^{1/2}=5\mu {\rm m}$. We also get
$\omega=15\rm{s}^{-1}$, $S=65 \rm{s}^{-1}$, and
$G/\hbar =0.2\rm{s}^{-1}\ll S$. We assume that the initial
condensate has a size of $(50 \mu\rm{m})^2$ which implies
$100$ occupied lattice sites with $n_l(0)\approx
10^7/100=10^5$ atoms per site and
$\sigma_l(0)\simeq[n_l(0)]^{1/2}\approx 300$ atoms per
site. Due to the measurement the dispersion shrinks $25$
times down to $\sigma_l(t=\infty) = (\omega
n_l(0)/S)^{1/4} \approx 12$ atoms per site in a timescale
$\tau_l=(S\omega n_l(0))^{-1/2}\approx 100 \mu\rm{s}$. 
Until now we have neglected the off-resonance Rayleigh scattering
due to the measurement. The same methods leading to 
Eqs.(\ref{LINDBLAD},\ref{KERNEL}) also predict a Rayleigh
depletion time of $\tau_D\approx 7\rm{ms}$, {\it i.e.} $70$ 
times longer than $\tau_l$ \cite{long}.

The dramatic $25$-fold shrinking of the dispersion in
occupation number results in a {\it quasi-}Fock state. An
{\it exact} Fock state would be 
$\hat{\Psi}^{\dagger}_{l_1}\dots \hat{\Psi}^{\dagger}_{l_N}|0\rangle$, 
where each atom can
be assigned to exactly one lattice site and as such it is
localized within the kernel resolution of $\Delta r^2$.  
The localization in an exact Fock state would cost a lot of
kinetic energy. In our {\it quasi-}Fock state, as shown by
the calculation of the expectation value of the hopping term in
our ansatz state (\ref{ANSATZ}), as long as all $\sigma_l\gg 1$
the energy remains very close to the energy of the initial
mean-field state.

A direct experimental test of our predictions and in particular of a 
simplified lattice model (\ref{LATTICE}) can be made in the 1-D array of 
weakly linked mesoscopic traps of the experiment described in 
Ref.\cite{orzel}.
In this experiment the traps are created by a standing optical
wave of wavelength $\lambda_{\rm st}=840 {\rm nm}$ superimposed on
a confining harmonic potential due to a magnetic trap. 
Given a transverse trapping angular frequency of $2 \pi \times 
120 \rm{s}^{1}$ we 
estimate the transverse length of the condensates in each trap as the 
corresponding oscillator length, equal
to $\xi=1 \mu{\rm m}$. In the experiment each
condensate has a $1/e$ width of $\Delta l=\lambda_{\rm st}/6$, 
therefore we estimate
$G/\hbar = g/( 2 \hbar \Delta l (2\xi)^2 ) \approx 40 \rm{s}^{-1}$. Assuming 
one is in the range of parameters of the experiment for which the
ground state of the condensate is close to a mean-field (quasi-coherent)
state (for instance squeezing $s \simeq 3$ in Figure 2 of Ref.\cite{orzel}), 
and that
 in each trap there are $n_l(0) \approx 100$ atoms 
of $^{87}{\rm Rb}$,
the hopping frequency is $\omega=G n_l(0)/\hbar s^2 \approx 500 \rm{s}^{-1}$.

\begin{figure}[t] 
\centering 
\epsfig{file=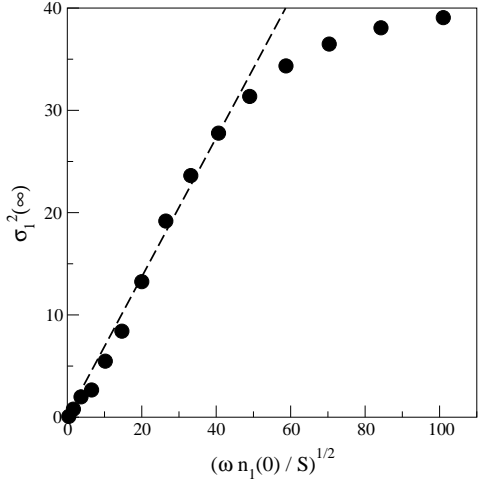, width=7cm, clip=}
\caption{Asymptotic variance $\sigma_1^2(\infty)$ versus 
$(\omega n_1(0)/S)^{1/2}$ for a 1-D 3-mode configuration, with $S=2 
\rm{s}^{-1}$ and $n_l(0)=100$. 
The scaling $\sigma_l^2(\infty)\sim\omega^{1/2}$ applies for 
$(\omega n_l(0)/S)^{1/2}\ll\sigma_l^2(0)=67$ when 
$\omega \ll S \sigma^4_l(0)/n_l$. The straight line is a guide 
for the eye. In the opposite case 
$\sigma_l^2(\infty)$ does not increase above the dispersion 
of the initial mean-field state. 
The $\sigma_l^2(\infty)$ are averages over the noisy values at 
large $t$ of trajectories in Fig.1. }
\label{figure2} 
\end{figure}     

Suppose now that we add phase contrast imaging to the above experimental 
setup. The kernel resolution is 
$\Delta r =(\pi \xi \lambda)^{1/2} \approx 1.5 \mu{\rm m}$, 
which implies there are 2 optical traps within
the kernel resolution. The effective number of atoms within $\Delta r$
is therefore $n_{\rm pci} =200$, the effective self coupling is 
$G_{\rm pci}/\hbar= G/2 \hbar=20 \rm{s}^{-1}$ and the effective hopping 
frequency is $\omega_{\rm pci} = \omega/2 = 250 \rm{s}^{-1}$. 
For a laser intensity of, say,
$I=100 {\rm mW}/{\rm cm}^2$, we obtain $S=500 \rm{s}^{-1}$ and 
$\sigma_l(\infty) = 3$ atoms per site
after a localization time of $\tau_l = 200 \mu{\rm s}$ but
much before a depletion time of $\tau_D=7 \rm{ms}$. The final
dispersion in the number of atoms per optical trap is 
$\sigma(\infty) = (\sigma_l(\infty)^2 / 2)^{1/2} = 2$. 
The final squeezing factor will be 
$\sigma^2(0)/\sigma^2(\infty) \approx 50$, where we
take $\sigma_l^2(0)=n_l(0)$. Such an enhancement of squeezing due to the 
measurement should be observable with the techniques of the experiment 
\cite{orzel}

In conclusion, we have applied the theory of open quantum systems to include
the back-action of a nondemolitive measurement into the dynamics of a 
Bose-Einstein
condensate.  We have shown that dispersive imaging with a
sufficiently high laser intensity results in number squeezing of the 
condensate state. Our prediction can be tested in present-day experiments
and could result in a significant improvement toward the 
implementation of Heisenberg-limited atom interferometry \cite{Bouyer}, 
and in a more general understanding of the back-action of other quantum-limited
devices \cite{nondemo}.

{\bf Acknowledgements.} We thank L. Viola for useful discussions. 
D.A.R.D. and J.D. were supported 
in part by NSA.

%\vspace*{-5mm} 

\end{document}